\documentclass[%
reprint,
 amsmath,amssymb,
 aps,
]{revtex4-1}

\usepackage{graphicx}
\usepackage{dcolumn}
\usepackage{bm}
\usepackage{xcolor}
\usepackage{verbatim}
\usepackage{chemformula}

\newcommand{\bmrm}[1]{\bm{\mathrm{#1}}}
\renewcommand{\vec}{\bmrm}

\begin{document}

\title{Using differentiable programming to obtain an energy and density-optimized exchange-correlation functional}
\author{Sebastian Dick$^{1,2}$}
\email{sebastian.dick@stonybrook.edu}
\address{Department of Physics and Astronomy
Stony Brook University
Stony Brook, NY 11794-3800}
\affiliation{%
$^1$Physics and Astronomy Department, Stony Brook University, Stony Brook, New York 11794-3800, United States
}%

\affiliation{%
$^2$Institute for Advanced Computational Science, Stony Brook University, Stony Brook, New York 11794-3800, United States
}%
\author{Marivi Fernandez-Serra$^{1,2}$}%
\email{maria.fernandez-serra@stonybrook.edu}
\address{Department of Physics and Astronomy
Stony Brook University
Stony Brook, NY 11794-3800}

\affiliation{%
$^1$Physics and Astronomy Department, Stony Brook University, Stony Brook, New York 11794-3800, United States
}%
\affiliation{%
$^2$Institute for Advanced Computational Science, Stony Brook University, Stony Brook, New York 11794-3800, United States
}%

\begin{abstract}
Using an end-to-end differentiable implementation of the Kohn-Sham self-consistent field equations, we obtain an accurate neural network-based exchange and correlation (XC) functional of the electronic density. The functional is optimized using information on both energy and density while exact constraints are enforced through an appropriate neural network architecture.
We evaluate our model against different families of XC approximations and show that, for non-empirical functionals, there is a strong linear correlation between energy and density errors.
Using this correlation, we define a novel XC functional quality metric that includes both energy and density errors, leading to a new, improved way to rank different approximations. Judged by this metric, our machine-learned functional significantly outperforms those within the same rung of approximations.
\end{abstract}
\maketitle

\preprint{APS/123-QED}

Density functional theory (DFT) serves without doubt as the workhorse method for electronic structure simulations in materials science and physics and has gained popularity within the chemistry community in recent decades.
This is in no small part due to its favorable scaling, allowing users to tackle systems sizes out of reach for most correlated wavefunction methods. 
However, inferences made from numerical simulations are only ever as good as their underlying approximations. This remains true for DFT, where these approximations are bundled somewhat opaquely in the elusive exchange-correlation (XC) functional. The Hohenberg-Kohn theorem guarantees that if this functional were known, ground-state properties of any interacting many-electron system could be described exactly \cite{hohenberg1964inhomogeneous}. In practice, one needs to pick from a plethora of different approximations, which often boils down to finding the right functional, cost and accuracy-wise, for the problem at hand.

It comes as no surprise that developing new and more accurate density functionals is a field of research on its own. Practitioners of this field generally have worked following two orthogonal approaches. Going back to Perdew and Wang \cite{perdew1992atoms}, one approach tries to develop functionals from first-principles only, without any empirically-fit parameters. Some of the most notable functionals from this family include PBE \cite{pbe}, TPSS \cite{tpss}, and SCAN \cite{sun2015strongly} which have proven themselves to be both versatile and reliable. Another approach, pioneered by Becke \cite{becke1988density}, is to fit functionals containing empirical parameters to either experimental or highly accurate simulated data. The size of these datasets can range from a few atoms to thousands of molecules and chemical reactions. These empirically fitted functionals such as B(3)LYP \cite{becke1988density, lee1988development, b3lyp} and more recently M06 \cite{m06l} and $\omega$B97X-V \cite{wb97xv} are often considered the gold standard for calculations in chemistry, but have so far failed to find widespread application in the solid-state community.

Beyond improving energies, approaching the exact functional should also lead to more accurate densities.
However, a recent study suggested that most empirically fitted functionals fall short of this expectation \cite{medvedev2017density}. 
This is concerning, not only from a theoretical point of view but also for practical reasons.
For example, the quality of a functional's electronic density is related to its ability to correctly describe a molecule's response to an external electric field \cite{hait2021too}.

A guided approach towards empirical functionals that produce better densities is clearly needed.
This task poses great challenges, as the Kohn-Sham equations introduce a non-linear relationship between functional form and self-consistent density.
A guided optimization of such functionals requires knowledge of the gradients  of the functional with respect to changes in the density.
Pioneering work by Nagai et al. \cite{nagai2019completing} circumvented the problem of missing gradients by adopting a Markov Chain Monte Carlo approach to optimize a functional.
DeepKS \cite{Chen2021} uses a stochastic term in its loss function that drives the functional towards the correct density.
A breakthrough solution to this problem was very recently proposed by Li et al. \cite{Li2021}, using differentiable programming.
By implementing the solution to the Kohn-Sham equations in JAX, a Python library that supports automatic differentiation on arbitrary operations, they can probe the electron density response to changes in the functional parametrization. 
The authors further showed that incorporating physical knowledge in the form of Kohn-Sham equations into the optimization algorithm has a regularizing effect making the algorithm more data-efficient.
Their work, however, was limited to the study of 1-d model systems. 

Here we optimize a density functional using an end-to-end differentiable implementation of the Kohn-Sham equations.
In contrast to other approaches that have used machine learning to approximate the exact functional \cite{dick2019learning,dick2020machine,Chen2021,nagai2019completing}, we impose a set of known constraint on the functional form.
These include the local Lieb-Oxford bound \cite{lieb1981improved} (LOB), which proves to be an important ingredient to obtaining
a more transferable model.

We aim at creating a general purpose functional,
capable of extrapolating well beyond its training
scope.
However, to make our training process computationally feasible, we mostly limit our training set to linear systems.
We show that this can be done without loss of generality, meaning that a thus optimized functional is still applicable to more complex molecules. 
With the goal of obtaining a model with a good balance
between computational cost and accuracy, we choose to optimize a neural network-based meta-GGA functional.
We demonstrate that it outperforms other meta-GGA functionals on a diverse selection of datasets both with respect to energy as well as density predictions.
Our analysis of different functionals identifies a high linear correlation between energy and density errors for non-empirical models.
Using this relationship we define a novel compound metric which we term energy-density error. 
Ranked by this compound error, our model is competitive even with functionals of the hybrid family.

At the heart of Kohn-Sham (KS) density functional theory lie the KS-equations

\begin{equation}
    \left\{ -\frac{1}{2}\nabla^2 + v_s[n](\vec{r}) \right\}\psi_i(\vec r) = \epsilon_i \psi_i(\vec r)\label{eq:ks}
\end{equation}

In this approach, the electron density $n(\vec r)$ is computed from the occupied one-particle orbitals 
$n(\vec r) = \sum_i^N |\psi_i(\vec r)|^2$, and the potential is given as
\begin{equation}
v_s[n](\vec r ) = v_{ext}(\vec r) + v_{H}[n](\vec r ) + v_{xc}[n,\vec \omega](\vec r ).
\end{equation}
Here, $v_{ext}(\vec r)$ is the external potential created by the ion cores, $v_{H}(\vec r)$ is the Hartree potential capturing the Coulomb interaction of the density with itself and $v_{xc}(\vec r)$ is the functional derivative of the  exchange-correlation functional with respect to the electron density $ v_{xc}(\vec r)= \frac{\delta E_{xc}[n, \vec \omega]}{\delta n(\vec r)}$. As all quantities except for the exchange-correlation functional are known, the goal of this work will be to find a parametrization $\vec \omega$ of $E_{xc}$ which accurately reproduces reference energies and electron densities while generalizing well to unseen systems.

As the potential depends on the density and therefore implicitly on the eigenstates $\psi_i$, the KS equations need to be solved iteratively. A popular Ansatz used in chemistry codes, and the one we choose here due to its efficiency for molecular systems, is to
expand the eigenstates in Eq.\ref{eq:ks} in terms of atom-centered Gaussian orbitals $\psi_i = \sum_{\mu}C_{i\mu}\phi_\mu$. 
One advantage of using a Gaussian basis is that integrals can be pre-computed analytically and stored to disk, reducing on-the-fly computations to simple tensor contractions. 
For this work, we have made use of the open-source python code PySCF \cite{sun2018pyscf, sun2020recent}.
We have re-implemented all routines needed to solve the Kohn-Sham equations to utilize PyTorch \cite{paszke2019pytorch}, making them end-to-end differentiable. 
One and two-electron integrals were computed with the original version of PySCF as the basis sets can be considered fixed for the purpose of this work.

A  fully differentiable implementation of the self-consistent field (SCF) method necessitates that gradients occurring for every mathematical operation, at every SCF iteration, be held in memory until they are used during back-propagation.
Especially tensor operations that involve grid points, such as the ones needed to generate the real-space density on which the xc-functional is evaluated, contribute a high memory cost. 
We have chosen to partially circumvent this problem by largely restricting our training set to linear closed-shell molecules during training.
We take advantage of their cylindrical symmetry by evaluating grid integrals on a reduced grid, namely a disk in the zx-plane.
To obtain the radial part of this grid, we make use of the methods provided by PySCF to generate Treutler-Ahlrichs type grids.
For the angular part, we use a simple Legendre-Gauss quadrature.
The size of the reduced grid is chosen so that it reproduces the number of electrons, integrated exchange-correlation energy as well as the exchange-correlation potential (in the atomic orbital basis) given by a reference calculation using a converged three-dimensional grid. 

\begin{figure*}[t!]
    \centering
    \includegraphics[width=.98\textwidth]{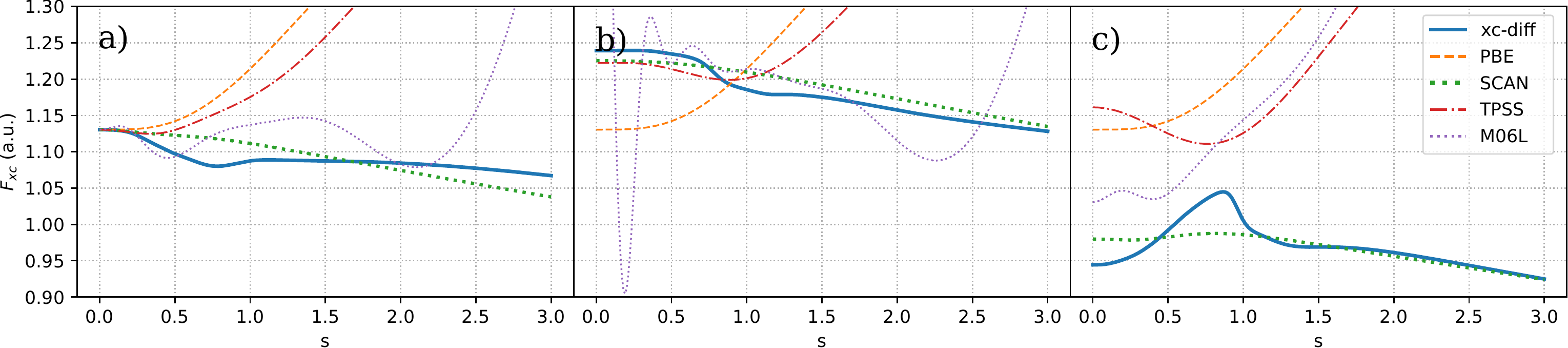}
    \caption{Exchange-correlation enhancement factors $F_{xc}$ for $r_s = 1$, $\zeta = 0$, and a) $\alpha = 1$, b) $\alpha = 0$, c) $\alpha = 10$}
    \label{fig:fxc}
\end{figure*}

We followed the common practice of defining the exchange correlation energy in terms of the energy per unit particle $E_{xc}[n,\vec \omega] =\int \epsilon_{xc}[n, \vec \omega](\vec r) n(\vec r ) d\vec r$.
We further decompose this energy density into its exchange and correlation parts  $\epsilon_{xc}[n, \vec \omega](\vec r) = \epsilon_{x}[n,\vec \omega_x](\vec r) + \epsilon_{c}[n,\vec \omega_c](\vec r)$ which are both independently parametrized. 
This allows us to factorize both functionals into fixed parts describing the behavior of a uniform electron gas (UEG) $e_{x/c}^{UEG}[n]$ and parametrized enhancement factors $F_{x/c}[n, \vec \omega_{x/c}]$ that take into account effects from inhomogeneities. 
The exchange  energy density of the UEG is given as $\epsilon_x[n](\vec r)= -\frac{3}{4} (3/\pi)^{1/3} n^{1/3}(\vec r)$, and the parametrization of $\epsilon_c$ by Perdew and Wang \cite{perdew1992accurate} was used. 
Rather than having our functionals depend on the electron density and its derivatives directly we define the following commonly used dimensionless quantities which will serve as input to our functionals:

\begin{align}
    x_0 &= n^{1/3} \\
    x_1 &= \frac{1}{2}\left\{(1+\zeta)^{4/3} + (1-\zeta)^{4/3}\right\} 
\end{align}
\begin{align}
    x_2 &= s =\frac{1}{2(3\pi^2)^{1/3}} \frac{|\nabla n|}{n^{4/3}}\\
    x_3 &= \alpha =\frac{\tau - \tau^W}{\tau^{unif}}
\end{align}
Where $\tau^W = |\nabla n|^2/8n$, $\tau^{unif} = (3/10)(3\pi^2)^{2/3}n^{5/3}$ and $\zeta$ corresponds to the spin polarization.

As neural networks struggle with handling features that range over multiple orders of magnitude, we further transform our input features $x_{0-3}$ by applying logarithmic transformations

\begin{align}
    \tilde x_0 &= \log(x_0 + \varepsilon_{\log})\\
    \tilde x_1 &= \log(x_1 + \varepsilon_{\log}) \\
    \tilde x_2 &= \left\{1 - \exp(-x_2^2)\right\}\log(x_2 + 1)\\
    \tilde x_3 &= \log\left\{(x_3 + 1)/2\right\}
\end{align}
with $\epsilon_{\log} = 10^{-5}$.
$\tilde x_2$ is designed so that its first derivative vanishes  at $x_2=0$. 
This poses a soft constraint on the enhancement factors $F_{x/c}$ to have the same property. 
We have found that doing so greatly improves convergence, especially for periodic systems. 
Similar reasoning was applied to $\tilde x_3$ where the employed transformation lead to better behaved functionals than the more obvious choice $\log(x_3 + \varepsilon_{\log})$.

Both $F_x$ and $F_c$ were parametrized by a fully connected neural network with three hidden layers with 16 nodes each. 
As activation function, we have used the Gaussian error linear unit (GELU) \cite{gelu}.
We will denote the mapping induced by this neural network as $\mathcal F(\cdot)$.

We modify our neural network models to fulfill certain constraints and scaling laws that are known about the exact functional.
To make $E_x$ behave correctly under uniform scaling of the electron density and obey the spin-scaling relation, we drop the variables $x_0$ and $x_1$ in $F_x$. 
We further introduce a transformation $I_a(x)$ that maps its input $x$ to a finite interval $[-1, a-1]$:
\begin{equation}
I_a(x) = \frac{a}{1+(a-1)\exp(-x)} -1
\end{equation}
while maintaining $I_a(0) = 0$. 
In the case of $F_x$, $I_{1.174}(x)$ is used to strictly enforce a local LOB\cite{lieb1981improved} $a=1.174$, whereas for $F_c$ we use $I_2(x)$ to ensure non-negativity of the enhancement factor.
Collecting all input features into a vector $\tilde{\vec x}$, the models can be written as:
\begin{align}
    &F_x(\tilde x_2,\tilde x_3) = 1 + I_{1.174}((\tilde x_2+\tanh^2{\tilde x_3})\mathcal F(\tilde x_2,\tilde x_3,\vec \omega_x))\\
    &F_c(\tilde{\vec x}) = 1 +  I_2((\tilde x_2+\tanh^2{\tilde x_3}) \mathcal F(\tilde{\vec x},\vec \omega_c))
\end{align}
 The factor $(\tilde x_2+\tanh^2{\tilde x_3})$ ensures that the UEG limit is recovered for $s=x_2=0$ ($\tilde x_2=0$) and $\alpha=x_3=1$ ($\tilde x_3=0$).

The datasets used in this work for training and validation consist of 21 atomization energies taken from the G2/97 set \cite{curtiss1997assessment},  three barrier heights taken from BH76 by Zhao et al \cite{zhao2005benchmark} and two reference ionization potentials from IP13 provided in \cite{lynch2003effectiveness}.
For the G2/97 dataset, we use atomization energies that were recalculated by Haunshild et al. \cite{haunschild2012new} and are considered more reliable than the enthalpies of formation given in the original version of the dataset. 

We augmented the G2/97 dataset with ground-state electron densities that we computed at the CCSD(T) level using the 6-311++G(3df,2pd) basis set, the same basis used for training the functionals.
Total atomic energies were taken from Ref. \citealp{chakravorty1993ground} and included in the training set as well.
Atomic electron densities were calculated and included for \ch{H} and \ch{Li}.
For model validation, during training, we used a disjoint subset of the data listed above, consisting of 8 atomization energies and densities from G2/97, and two reference barrier heights from BH76. 
A detailed list of the structures used for training and validation can be found in the SI.

Models were pre-trained to match SCAN \cite{sun2015strongly}  on the 21 molecules contained in the training set by randomly sampling the exchange enhancement factor on molecular grids and fitting to it.
The functional parameters are then trained to optimize a compound loss, combining errors in total energies $E^{(i)}_{j;tot}$ and reaction energies (which includes atomization energies and barrier heights) $E^{(i)}_{j; RE}$ at SCF iteration $j$, as well as electron densities $n^{(i)}$.
\begin{align}
    \mathcal{L} &= \lambda_E\mathcal{L}_E + \lambda_{RE}\mathcal{L}_{RE} + \lambda_{n}\mathcal{L}_{n}\\
    \mathcal{L}_E &= \mathbb{E}\left[ \sum_{j=10}^{25} \left\{w_j (E^{(i)}_{j;tot,ref} - E^{(i)}_{j;tot})\right\}^2\right]\\
        \mathcal{L}_{RE} &= \mathbb{E}\left[ \sum_{j=10}^{25} \left\{w_j (E^{(i)}_{j;RE,ref} - E^{(i)}_{j;RE})\right\}^2\right]
\end{align}
\begin{align}
    \mathcal{L}_n &= \mathbb{E}\left[ l^{(i)}_n\right]\\
    l^{(i)}_n &= \frac{1}{N_e^2} \int_{\vec{r}} (n^{(i)}(\vec r) - n^{(i)}_{ref}(\vec r))^2 \label{eq:rholoss}
\end{align}
with flexible weights $\lambda_E$, $\lambda_{RE}$, $\lambda_n$ and expectation values taken over the training set. 
We set the weights to $\lambda_{RE} = 1$, $\lambda_n = 20$, $\lambda_E = 0.01$. 
Rather than including only converged energies in our loss function, we follow the approach by Li et. al \cite{Li2021}, employing $w_j = \left(\frac{j-10}{15}\right)^2$ that penalize solutions which lead to slowly converging SCF calculations.

The functional parameters are optimized using Adam with an initial learning rate $10^{-4}$  which is decayed by a factor of $0.1$ after every ten consecutive epochs without a decrease in training loss. We employ an $l_2$-regularization of $10^{-6}$ and a batch size of one reaction.

\begin{figure}[t!]
    \centering
    \includegraphics[width=.48\textwidth]{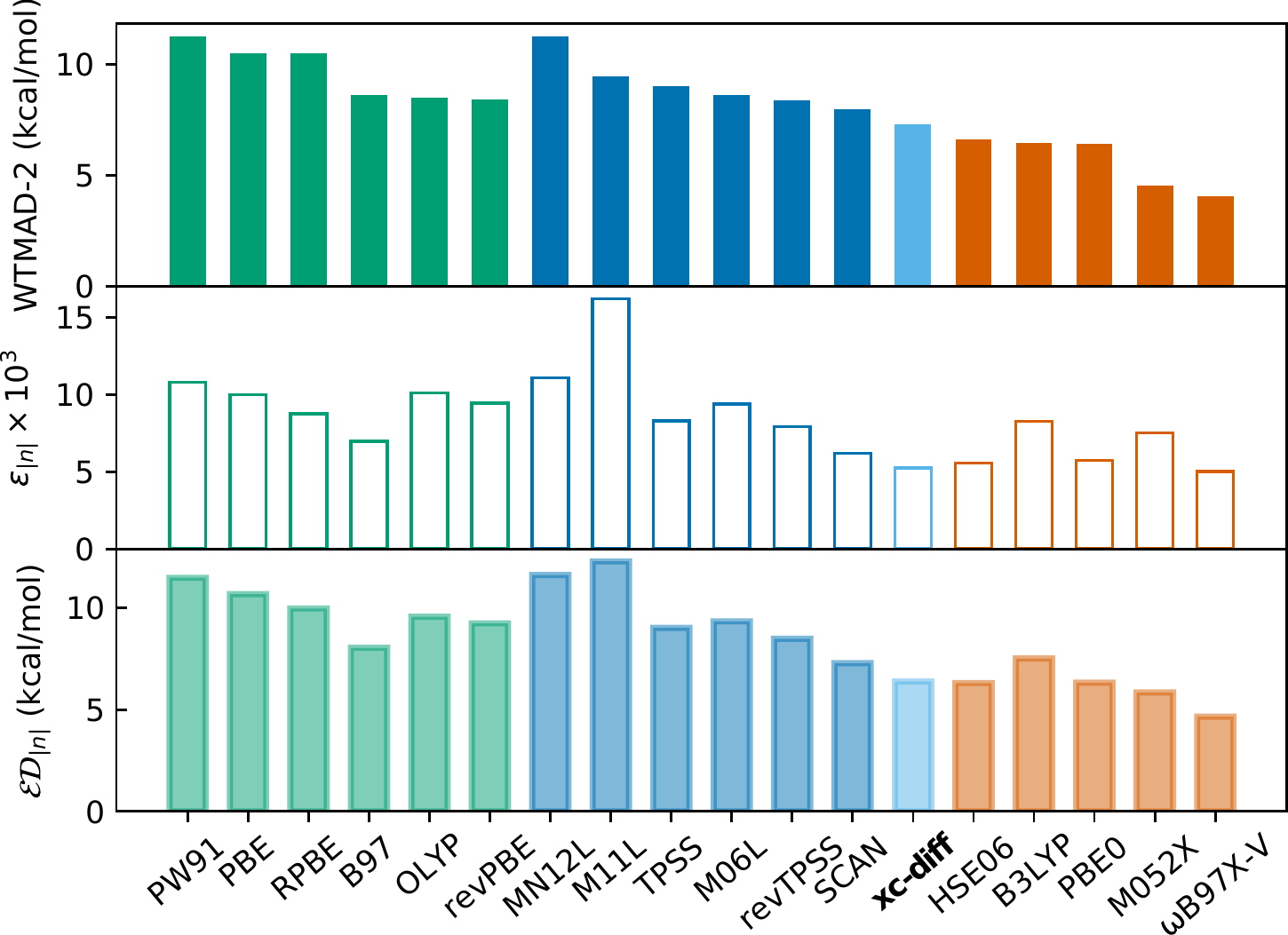}
    \caption{Weighted mean absolute deviations (WTMAD-2) (top), density errors (center) and energy-density error (bottom) for several functionals including our model, xc-diff.}
    \label{fig:wtmad2}
\end{figure}

We tested our functional on 140 atomization energies contained in the W4-11 \cite{w4-11} dataset, 76 barrier heights of hydrogen transfer, heavy atom transfer, nucleophilic substitution, unimolecular and association reactions from BH76, and 43 decomposition energies of artificial molecules contained in the MB16-43 \cite{goerigk2017look} dataset.
To achieve a wider assessment of our functional we further tested it on the diverse diet-GMTKN55 dataset \cite{gould2018diet}. 
GMTKN55 consists of 55 subsets that each probe different properties of a given functional. 
The subsets can be divided into categories by interaction type. 
These categories comprise reaction energies for small systems, reaction energies for large systems and isomerization reactions, barrier heights, intermolecular noncovalent interactions, and intramolecular noncovalent interactions. 
Diet-GMTKN55 provides representative sub-samples of GMTKN55 that have been shown to lead to the same ranking of DFs as the full dataset, at a significantly reduced computational cost.

We choose to evaluate our functional on the proposed 150 samples, the largest 'diet' dataset,
using a weighted mean of mean absolute deviations (MAD) across the subsets.  
The weights are chosen by Gould to reproduce the WTMAD-2 weighted mean of means proposed by Goerig et al., which scales the mean absolute energy deviations $\text{MAD}_i$ of a subset $i$ containing $N_i$ reactions by the inverse energy range of a given subset $\overline{|\Delta E|}_i$
\begin{equation}
    \text{WTMAD-2} = N^{-1} \sum_i^{55} N_i \cdot \frac{ 56.84 \text{kcal mol}^{-1} }{\overline{|\Delta E|}_i}\cdot \text{MAD}_i,
\end{equation}
with $N = \sum_i^{55} N_i$.
 The goal is to give more weight to datasets with little variation in the energy and to scale down systems with large variations. 

We conducted all necessary single-point calculations with PySCF using our in-house code libnxc \cite{libnxc} as a plug-in to allow for the use of PyTorch xc-models. We employed the  def2-QZVP basis set and augmented it with diffuse functions for the subsets recommended in Ref. \citealp{goerigk2017look}. A PySCF grid level of 3 together with an energy convergence tolerance of $10^{-8} E_h$ was chosen.

\begin{table}
\begin{tabular}{r|ccc|ccc}
\toprule
{} &  AE &  BH & DE &  WTMAD-2 &  $\varepsilon_{|n|} \times 10^3$ &  $\mathcal E \mathcal D_{|n|}$ \\
\hline
RPBE \cite{rpbe}    &    8.3 &   9.0 &     50.8 &     10.5 &     8.8 &      10.0 \\
B97 \cite{b97}    &    4.7 &   7.3 &     36.1 &      8.6 &     7.0 &       8.0 \\
OLYP \cite{olyp}   &    9.9 &   8.5 &     29.0 &      8.5 &    10.1 &       9.6 \\
revPBE \cite{revPBE}  &    7.6 &   8.3 &     27.1 &      8.4 &     9.4 &       9.2 \\
\hline
M06L    &    4.4 &   3.9 &     63.3 &      8.6 &     9.4 &       9.3 \\
revTPSS &    5.7 &   8.9 &     36.7 &      8.4 &     7.9 &       8.5 \\
SCAN    &    4.1 &   7.8 &     17.8 &      8.0 &     6.2 &       7.3 \\
\textbf{xc-diff} &    3.5 &   6.5 &     22.7 &      7.3 &     5.2 &       6.4 \\
\hline
PBE0    &    3.7 &   5.0 &     15.9 &      6.4 &     5.7 &       6.3 \\
B3LYP   &    3.4 &   5.7 &     24.8 &      6.5 &     8.3 &       7.5 \\
M05-2X \cite{m052x}  &    4.0 &   1.7 &     26.3 &      4.6 &     7.5 &       5.8 \\
$\omega$B97X-V \cite{wb97xv} &    2.8 &   1.8 &     32.5 &      4.1 &     5.0 &       4.7 \\
\hline
\end{tabular}
\caption{Mean absolute errors in kcal mol$^{-1}$ for atomization energies (AE) over the W4-11 dataset, barrier heights (BH) in BH76 and decomposition energies (DE) for MB16-43. Weighted means WTMAD-2 and $\Delta$ are also given in kcal mol$^{-1}$. Mean density error $\varepsilon_n$ is unit-less. A complete list of functionals is provided in the SI.} \label{tab:errors}
\end{table}
To ensure the correct treatment of non-covalent interactions, all results reported include the DFT-D3 dispersion correction with Becke-Johnson damping \cite{grimme2010consistent, grimme2011effect}. Parameters for our functional were optimized following the procedure outlined in Ref. \citealp{goerigk2017look} and are summarized in the SI.

Fig.\ref{fig:fxc} shows a comparison of XC-enhancement factors $F_\text {xc} = \varepsilon_\text {xc}/\varepsilon_\text {x,UEG}$ for a set of density functionals. Our functional, which we call xc-diff, resembles TPSS \cite{tpss} and SCAN for $\alpha=1$ and small $s$. In many instances, xc-diff qualitatively resembles SCAN which is most likely due to the chosen initialization procedure as well as the imposed local LOB of 1.174. In the weakly interacting regime (large $\alpha$) xc-diff seems closer in behavior to M06L \cite{m06l} and TPSS for small to moderate s while reverting to the correct decay given by SCAN for large s. Despite the small regularization employed, the obtained neural network-based functional is smooth and no problems regarding convergence during SCF calculations were encountered. We accredit this to the optimization procedure and the weighted loss which penalized parametrizations that would lead to slowly converging calculations.  

Comparing the weighted means WTMAD-2 shown in Fig. \ref{fig:wtmad2} and Tab. \ref{tab:errors}, we see that xc-diff outperforms  SCAN, (rev)TPSS \cite{revTPSS} , and the empirically fitted Minnesota functionals M06L, M11L \cite{m11l} and MN12L \cite{mn12l}. 
It should be pointed out that the training sets used to optimize the Minnesota functionals were about one order of magnitude larger than the one used in this work. 

The datasets W4-11, BH76, and MB16-43  illuminate the strengths and weaknesses of the respective functionals.
For atomization energies of small systems, xc-diff outperforms SCAN by 0.6 kcal mol$^{-1}$ and is comparable to the global hybrids B3LYP \cite{b3lyp} and PBE0 \cite{pbe0}.
Being susceptible to delocalization errors, barrier heights pose a challenge to semi-local functionals. Here,  xc-diff outperforms SCAN by more than 1 kcal mol$^{-1}$ but is outperformed by about the same amount by  PBE0 and B3LYP.
Not fully shown in Tab. \ref{tab:errors} due to their large WTMAD-2, the Minnesota functionals provide an excellent treatment of this dataset with MAEs ranging from 3.9 to 1.7  kcal mol$^{-1}$.
However, it is worth noting that barrier heights played a major role in the training sets used to optimize all Minnesota functionals, so their accuracy comes as no surprise. 
MB14-36 plays a special role as it contains artificial, randomly generated molecules and has proven challenging especially to empirical functionals. Here, xc-diff is less accurate than SCAN but shows reasonable performance compared to all other functionals considered here.

\begin{figure}[t!]
    \centering
    \includegraphics[width=.48\textwidth]{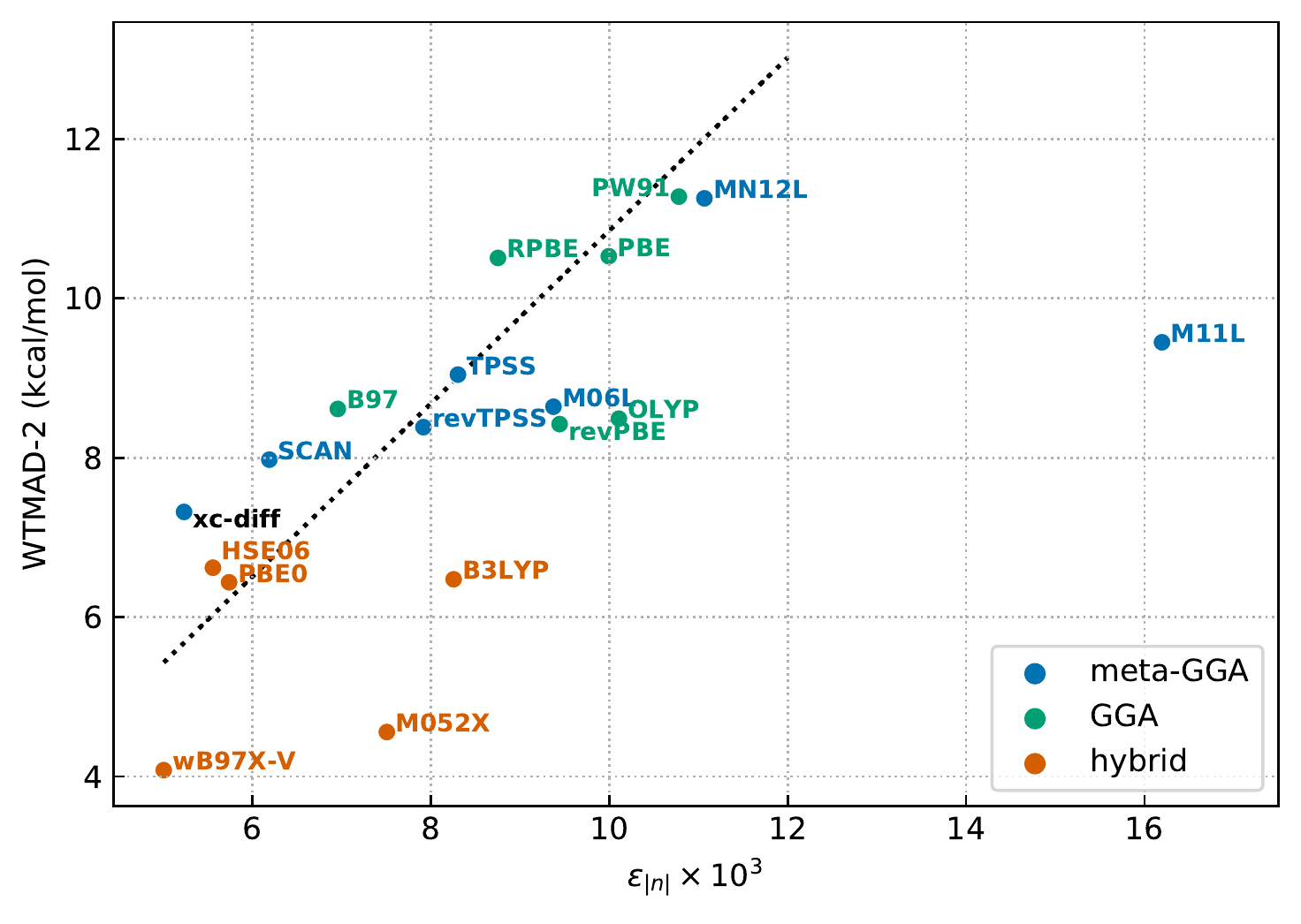}
    \caption{Correlation plot for density error and total WTMAD-2. Dotted line indicates best linear fit to non-empirical DFs }
    \label{fig:corr} 
\end{figure}

Beyond comparing energies, we used the previously calculated CCDS(T) electron densities across the G2/97 dataset to assess the accuracy of our functional regarding densities. Mean errors across the dataset were computed using the metric
\begin{equation}
    \varepsilon_{|n|} = \mathbb{E} \left[\frac{1}{N_e}\int_{\vec{r}} |n^{(i)}(\vec r) - n^{(i)}_{ref}(\vec r)|\right]\label{eq:densityerror}
\end{equation}
The methods were identical to those used for the diet-GMTKN55 dataset except for the basis set, which was chosen as 6-311++G(3df,2pd) for easier comparison with our coupled-cluster reference densities.

Fig. \ref{fig:wtmad2}  shows that xc-diff outperforms all other tested meta-GGA functionals by a clear margin. We further included data obtained with global hybrids and GGAs. While most hybrids improve upon traditional meta-GGA functionals, xc-diff is still 9\% more accurate regarding the density than PBE0.

We believe that a functional should be judged by both its accuracy regarding energies as well as densities. A metric combining both energy and density errors would therefore be useful to score and rank functionals, however finding such a metric is no straightforward task. 

An important clue might be provided by the high linear correlation ($R^2 = 0.87$) between WTMAD-2 and density errors for non-empirical DFs (PW91\cite{perdew1992atoms}, PBE\cite{pbe}, TPSS\cite{tpss}, revTPSS\cite{revTPSS}, SCAN, PBE0\cite{pbe0}). The best fit of a linear regression model (with zero intercept) is shown in Fig. \ref{fig:corr} by a dotted line. Remarkably, regardless of the level of approximation, non-empirical fuctionals closely follow this trend-line, while many empirically fitted DFs seem to deviate significantly from it. We have been able to confirm that this trend persists for other definitions of the density error, such as one based on the Kullback-Leibler divergence (see SI for details). Our functional, xc-diff, shows a density error that is lower than expected from this trend.  

Inspired by this finding, we propose a new metric $\mathcal E \mathcal D$ that allows us to combine density with energy errors:

\begin{equation}
   \mathcal E \mathcal D_{|n|} = 2\left(\frac{1}{\text{WTMAD-2}} + \frac{1}{f_E(\varepsilon_{|n|})}\right)^{-1}.
\end{equation}

$f_E(\varepsilon_{|n|}) = \gamma \cdot \varepsilon_{|n|} $ with $\gamma = 1084.87 \text{ kcal mol}^{-1}$ corresponds to the linear regression model used in Fig. \ref{fig:corr}, and can be interpreted as the energy error (WTMAD-2) a fictional non-empirical functional with density error $\varepsilon_n$ would exhibit according to our model. 
Fig. \ref{fig:wtmad2} shows $\mathcal E \mathcal D_{|n|}$ across density functionals. We see that within meta-GGAs, the order of functionals remains largely unchanged but due to xc-diff's accuracy for densities, it now outperforms B3LYP
and matches the accuracy range of other popular hybrids such as PBE0.

Using an end-to-end differentiable implementation of the Kohn-Sham equations we have successfully optimized a highly accurate meta-GGA XC functional. 
Our results indicate that, despite exhaustive efforts in recent years to create better meta-GGA models, there is still room for improvement.  
It has been argued that such improvement should be achieved in a non-empirical approach imposing physically motivated exact constraints with a minimal number of free parameters \cite{perdew2005prescription}.
We have shown that a data-driven search using machine learning combined with an adherence to constraints can provide an equally valid path. 
A crucial ingredient of our method is given by automatic differentiation, which allows the optimization algorithm to make use of valuable information contained in the electron density, effectively enlarging the training set size. 
It remains to be tested how a thus optimized functional performs for solid systems; work that will be the subject of future research. 
While we believe that our functional could be further improved by fitting to larger training sets, its accuracy is inherently limited by the functional form of meta-GGAs.
We predict that advances in hardware development along with more efficient implementations of our code will soon allow us to apply our method to much larger training sets and higher rungs of DFT's Jacob's ladder \cite{perdew2005prescription}, opening the path towards  functionals of unprecedented accuracy.

\bibliographystyle{unsrt}
\bibliography{bibliography}

\begin{thebibliography}{10}

\bibitem{hohenberg1964inhomogeneous}
Pierre Hohenberg and Walter Kohn.
\newblock Inhomogeneous electron gas.
\newblock {\em Physical review}, 136(3B):B864, 1964.

\bibitem{perdew1992atoms}
John~P Perdew, John~A Chevary, Sy~H Vosko, Koblar~A Jackson, Mark~R Pederson,
  Dig~J Singh, and Carlos Fiolhais.
\newblock Atoms, molecules, solids, and surfaces: Applications of the
  generalized gradient approximation for exchange and correlation.
\newblock {\em Physical review B}, 46(11):6671, 1992.

\bibitem{pbe}
John~P Perdew, Kieron Burke, and Matthias Ernzerhof.
\newblock Generalized gradient approximation made simple.
\newblock {\em Physical review letters}, 77(18):3865, 1996.

\bibitem{tpss}
Jianmin Tao, John~P Perdew, Viktor~N Staroverov, and Gustavo~E Scuseria.
\newblock Climbing the density functional ladder: Nonempirical
  meta--generalized gradient approximation designed for molecules and solids.
\newblock {\em Physical Review Letters}, 91(14):146401, 2003.

\bibitem{sun2015strongly}
Jianwei Sun, Adrienn Ruzsinszky, and John~P Perdew.
\newblock Strongly constrained and appropriately normed semilocal density
  functional.
\newblock {\em Physical review letters}, 115(3):036402, 2015.

\bibitem{becke1988density}
Axel~D Becke.
\newblock Density-functional exchange-energy approximation with correct
  asymptotic behavior.
\newblock {\em Physical review A}, 38(6):3098, 1988.

\bibitem{lee1988development}
Chengteh Lee, Weitao Yang, and Robert~G Parr.
\newblock Development of the colle-salvetti correlation-energy formula into a
  functional of the electron density.
\newblock {\em Physical review B}, 37(2):785, 1988.

\bibitem{b3lyp}
Philip~J Stephens, Frank~J Devlin, Cary~F Chabalowski, and Michael~J Frisch.
\newblock Ab initio calculation of vibrational absorption and circular
  dichroism spectra using density functional force fields.
\newblock {\em The Journal of physical chemistry}, 98(45):11623--11627, 1994.

\bibitem{m06l}
Yan Zhao and Donald~G Truhlar.
\newblock A new local density functional for main-group thermochemistry,
  transition metal bonding, thermochemical kinetics, and noncovalent
  interactions.
\newblock {\em The Journal of chemical physics}, 125(19):194101, 2006.

\bibitem{wb97xv}
Narbe Mardirossian and Martin Head-Gordon.
\newblock $\omega$b97x-v: A 10-parameter, range-separated hybrid, generalized
  gradient approximation density functional with nonlocal correlation, designed
  by a survival-of-the-fittest strategy.
\newblock {\em Physical Chemistry Chemical Physics}, 16(21):9904--9924, 2014.

\bibitem{medvedev2017density}
Michael~G Medvedev, Ivan~S Bushmarinov, Jianwei Sun, John~P Perdew, and
  Konstantin~A Lyssenko.
\newblock Density functional theory is straying from the path toward the exact
  functional.
\newblock {\em Science}, 355(6320):49--52, 2017.

\bibitem{hait2021too}
Diptarka Hait, Yu~Hsuan Liang, and Martin Head-Gordon.
\newblock Too big, too small, or just right? a benchmark assessment of density
  functional theory for predicting the spatial extent of the electron density
  of small chemical systems.
\newblock {\em The Journal of chemical physics}, 154(7):074109, 2021.

\bibitem{nagai2019completing}
Ryo Nagai, Ryosuke Akashi, and Osamu Sugino.
\newblock {Completing density functional theory by machine-learning hidden
  messages from molecules}.
\newblock {\em arXiv preprint arXiv:1903.00238}, 2019.

\bibitem{Chen2021}
Yixiao Chen, Linfeng Zhang, Han Wang, and E.~Weinan.
\newblock {DeePKS: A Comprehensive Data-Driven Approach toward Chemically
  Accurate Density Functional Theory}.
\newblock {\em Journal of Chemical Theory and Computation}, 17(1):170--181,
  2021.

\bibitem{Li2021}
Li~Li, Stephan Hoyer, Ryan Pederson, Ruoxi Sun, Ekin~D. Cubuk, Patrick Riley,
  and Kieron Burke.
\newblock {Kohn-Sham Equations as Regularizer: Building Prior Knowledge into
  Machine-Learned Physics}.
\newblock {\em Physical Review Letters}, 126(3):1--7, 2021.

\bibitem{dick2019learning}
Sebastian Dick and Marivi Fernandez-Serra.
\newblock Learning from the density to correct total energy and forces in first
  principle simulations.
\newblock {\em The Journal of chemical physics}, 151(14):144102, 2019.

\bibitem{dick2020machine}
Sebastian Dick and Marivi Fernandez-Serra.
\newblock Machine learning accurate exchange and correlation functionals of the
  electronic density.
\newblock {\em Nature communications}, 11(1):1--10, 2020.

\bibitem{lieb1981improved}
Elliott~H Lieb and Stephen Oxford.
\newblock Improved lower bound on the indirect coulomb energy.
\newblock {\em International Journal of Quantum Chemistry}, 19(3):427--439,
  1981.

\bibitem{sun2018pyscf}
Qiming Sun, Timothy~C Berkelbach, Nick~S Blunt, George~H Booth, Sheng Guo,
  Zhendong Li, Junzi Liu, James~D McClain, Elvira~R Sayfutyarova, Sandeep
  Sharma, et~al.
\newblock Pyscf: the python-based simulations of chemistry framework.
\newblock {\em Wiley Interdisciplinary Reviews: Computational Molecular
  Science}, 8(1):e1340, 2018.

\bibitem{sun2020recent}
Qiming Sun, Xing Zhang, Samragni Banerjee, Peng Bao, Marc Barbry, Nick~S Blunt,
  Nikolay~A Bogdanov, George~H Booth, Jia Chen, Zhi-Hao Cui, et~al.
\newblock Recent developments in the pyscf program package.
\newblock {\em The Journal of chemical physics}, 153(2):024109, 2020.

\bibitem{paszke2019pytorch}
Adam Paszke, Sam Gross, Francisco Massa, Adam Lerer, James Bradbury, Gregory
  Chanan, Trevor Killeen, Zeming Lin, Natalia Gimelshein, Luca Antiga, et~al.
\newblock Pytorch: An imperative style, high-performance deep learning library.
\newblock {\em arXiv preprint arXiv:1912.01703}, 2019.

\bibitem{perdew1992accurate}
John~P Perdew and Yue Wang.
\newblock Accurate and simple analytic representation of the electron-gas
  correlation energy.
\newblock {\em Physical review B}, 45(23):13244, 1992.

\bibitem{gelu}
Dan Hendrycks and Kevin Gimpel.
\newblock Gaussian error linear units (gelus).
\newblock {\em arXiv preprint arXiv:1606.08415}, 2016.

\bibitem{curtiss1997assessment}
Larry~A Curtiss, Krishnan Raghavachari, Paul~C Redfern, and John~A Pople.
\newblock Assessment of gaussian-2 and density functional theories for the
  computation of enthalpies of formation.
\newblock {\em The Journal of Chemical Physics}, 106(3):1063--1079, 1997.

\bibitem{zhao2005benchmark}
Yan Zhao, N{\'u}ria Gonz{\'a}lez-Garc{\'\i}a, and Donald~G Truhlar.
\newblock Benchmark database of barrier heights for heavy atom transfer,
  nucleophilic substitution, association, and unimolecular reactions and its
  use to test theoretical methods.
\newblock {\em The Journal of Physical Chemistry A}, 109(9):2012--2018, 2005.

\bibitem{lynch2003effectiveness}
Benjamin~J Lynch, Yan Zhao, and Donald~G Truhlar.
\newblock Effectiveness of diffuse basis functions for calculating relative
  energies by density functional theory.
\newblock {\em The Journal of Physical Chemistry A}, 107(9):1384--1388, 2003.

\bibitem{haunschild2012new}
Robin Haunschild and Wim Klopper.
\newblock New accurate reference energies for the g2/97 test set.
\newblock {\em The Journal of chemical physics}, 136(16):164102, 2012.

\bibitem{chakravorty1993ground}
Subhas~J Chakravorty, Steven~R Gwaltney, Ernest~R Davidson, Farid~A Parpia, and
  Charlotte~Froese p~Fischer.
\newblock Ground-state correlation energies for atomic ions with 3 to 18
  electrons.
\newblock {\em Physical Review A}, 47(5):3649, 1993.

\bibitem{w4-11}
Amir Karton, Shauli Daon, and Jan~ML Martin.
\newblock W4-11: A high-confidence benchmark dataset for computational
  thermochemistry derived from first-principles w4 data.
\newblock {\em Chemical Physics Letters}, 510(4-6):165--178, 2011.

\bibitem{goerigk2017look}
Lars Goerigk, Andreas Hansen, Christoph Bauer, Stephan Ehrlich, Asim Najibi,
  and Stefan Grimme.
\newblock A look at the density functional theory zoo with the advanced gmtkn55
  database for general main group thermochemistry, kinetics and noncovalent
  interactions.
\newblock {\em Physical Chemistry Chemical Physics}, 19(48):32184--32215, 2017.

\bibitem{gould2018diet}
Tim Gould.
\newblock ‘diet gmtkn55’offers accelerated benchmarking through a
  representative subset approach.
\newblock {\em Physical Chemistry Chemical Physics}, 20(44):27735--27739, 2018.

\bibitem{libnxc}
Sebastian Dick.
\newblock libnxc.
\newblock \url{https://github.com/semodi/libnxc}, 2021.

\bibitem{rpbe}
BHLB Hammer, Lars~Bruno Hansen, and Jens~Kehlet N{\o}rskov.
\newblock Improved adsorption energetics within density-functional theory using
  revised perdew-burke-ernzerhof functionals.
\newblock {\em Physical review B}, 59(11):7413, 1999.

\bibitem{b97}
Axel~D Becke.
\newblock Density-functional thermochemistry. v. systematic optimization of
  exchange-correlation functionals.
\newblock {\em The Journal of chemical physics}, 107(20):8554--8560, 1997.

\bibitem{olyp}
Nicholas~C Handy and Aron~J Cohen.
\newblock Left-right correlation energy.
\newblock {\em Molecular Physics}, 99(5):403--412, 2001.

\bibitem{revPBE}
Yingkai Zhang and Weitao Yang.
\newblock Comment on “generalized gradient approximation made simple”.
\newblock {\em Physical Review Letters}, 80(4):890, 1998.

\bibitem{m052x}
Yan Zhao, Nathan~E Schultz, and Donald~G Truhlar.
\newblock Design of density functionals by combining the method of constraint
  satisfaction with parametrization for thermochemistry, thermochemical
  kinetics, and noncovalent interactions.
\newblock {\em Journal of chemical theory and computation}, 2(2):364--382,
  2006.

\bibitem{grimme2010consistent}
Stefan Grimme, Jens Antony, Stephan Ehrlich, and Helge Krieg.
\newblock A consistent and accurate ab initio parametrization of density
  functional dispersion correction (dft-d) for the 94 elements h-pu.
\newblock {\em The Journal of chemical physics}, 132(15):154104, 2010.

\bibitem{grimme2011effect}
Stefan Grimme, Stephan Ehrlich, and Lars Goerigk.
\newblock Effect of the damping function in dispersion corrected density
  functional theory.
\newblock {\em Journal of computational chemistry}, 32(7):1456--1465, 2011.

\bibitem{revTPSS}
John~P Perdew, Adrienn Ruzsinszky, G{\'a}bor~I Csonka, Lucian~A Constantin, and
  Jianwei Sun.
\newblock Workhorse semilocal density functional for condensed matter physics
  and quantum chemistry.
\newblock {\em Physical Review Letters}, 103(2):026403, 2009.

\bibitem{m11l}
Roberto Peverati and Donald~G Truhlar.
\newblock M11-l: A local density functional that provides improved accuracy for
  electronic structure calculations in chemistry and physics.
\newblock {\em The Journal of Physical Chemistry Letters}, 3(1):117--124, 2012.

\bibitem{mn12l}
Roberto Peverati and Donald~G Truhlar.
\newblock An improved and broadly accurate local approximation to the
  exchange--correlation density functional: The mn12-l functional for
  electronic structure calculations in chemistry and physics.
\newblock {\em Physical Chemistry Chemical Physics}, 14(38):13171--13174, 2012.

\bibitem{pbe0}
Carlo Adamo and Vincenzo Barone.
\newblock Toward reliable density functional methods without adjustable
  parameters: The pbe0 model.
\newblock {\em The Journal of chemical physics}, 110(13):6158--6170, 1999.

\bibitem{perdew2005prescription}
John~P Perdew, Adrienn Ruzsinszky, Jianmin Tao, Viktor~N Staroverov, Gustavo~E
  Scuseria, and G{\'a}bor~I Csonka.
\newblock Prescription for the design and selection of density functional
  approximations: More constraint satisfaction with fewer fits.
\newblock {\em The Journal of chemical physics}, 123(6):062201, 2005.

\bibitem{kullback1951information}
Solomon Kullback and Richard~A Leibler.
\newblock On information and sufficiency.
\newblock {\em The annals of mathematical statistics}, 22(1):79--86, 1951.

\bibitem{brandenburg2016benchmark}
JG~Brandenburg, JE~Bates, J~Sun, and JP~Perdew.
\newblock Benchmark tests of a strongly constrained semilocal functional with a
  long-range dispersion correction.
\newblock {\em Physical Review B}, 94(11):115144, 2016.

\bibitem{hse06}
Aliaksandr~V Krukau, Oleg~A Vydrov, Artur~F Izmaylov, and Gustavo~E Scuseria.
\newblock Influence of the exchange screening parameter on the performance of
  screened hybrid functionals.
\newblock {\em The Journal of chemical physics}, 125(22):224106, 2006.

\end{thebibliography}

\section{Supplementary Information}

\subsection{Restoring the 3-d potential}

While it is straightforward to obtain the correct xc-energy with the reduced (symmetrized) grid, caution needs to be exercised when calculating the potential. Using the automatic differentiation capabilities of our SCF implementation, the xc-potential $V_{xc,\mu\nu}$ in the atomic orbital (AO) basis can be obtained with
\begin{equation}
V_{xc,\mu\nu} \equiv \int \frac{\delta E_{xc}}{\delta n(\vec r)} \phi_{\mu}(\vec r) \phi_{\nu}(\vec r) d^3 r = \frac{\partial E_{xc}}{\partial \rho_{\mu\nu}}\label{eq:potential}
\end{equation}
where $\rho_{\mu\nu}$ is the density matrix in the AO basis. 
If $E_{xc}$ was merely evaluated on the reduced grid, taking the partial derivative effectively neglects the integration over the azimuthal angle $\phi$ in Eq. \ref{eq:potential}. If $\tilde V_{xc,\mu\nu}$ is the thus obtained (incorrect) potential, the correct potential can be restored by first symmetrizing $\tilde V_{xc,\mu\nu}$ over angular momentum projections of opposite sign $( m \leftrightarrow  -m)$, followed by an element-wise multiplication with a boolean mask ensuring that the relation
\begin{equation}
    \int \phi_{\mu}(\vec r) \phi_{\nu}(\vec r) d\phi \propto \delta_{m_\mu m_\nu}.
\end{equation} 
is recovered. In this notation $m_\mu$,$m_\nu$ correspond to the angular momentum projections associated with the AOs $\mu$ and $\nu$.

\subsection{Training data}

We train the functional on a subset of G2/97 \cite{curtiss1997assessment} consisting of 10 linear, closed shell molecules: \ch{H2}, \ch{N2}, \ch{LiF} , \ch{CNH}, \ch{CO2}, \ch{F2}, \ch{C2H2}, \ch{OC}, \ch{LiH}, \ch{Na2} and
3 linear open shell molecules: \ch{NO}, \ch{CH}, \ch{OH}.
All closed-shell linear molecules, as well as \ch{NO}, were calculated on a symmetrized grid as described above. We restored the cylindrical symmetry of the NO radical charge density by enforcing an artificial spin quartet state, which allowed us to use the symmetrization procedure described above. \ch{CH} and \ch{OH} were computed on a full three-dimensional grid. 

We found it beneficial to include an additional set of 8 non-linear closed shell molecules: \ch{NO2}, \ch{NH}, \ch{O3}, \ch{N2O}, \ch{CH3}, \ch{CH2}, \ch{H_2O}, \ch{NH3}. We further added the ionization potential of \ch{Li} and \ch{C} along with barrier heights for the following reactions taken from BH76 \cite{zhao2005benchmark}: \ch{OH} + \ch{N2} $\rightarrow$ \ch{H} + \ch{N2O}, \ch{OH} + \ch{CH3} $\rightarrow$ \ch{O} + \ch{CH4}, \ch{HF} + \ch{F} $\rightarrow$ \ch{H} + \ch{F2}. Due to the significant computational cost of our algorithm for large grids we treated these reactions and non-linear molecules non-selfconsistently during training, using a converged electron density obtained with SCAN as input. For these systems, we scaled $\lambda_{RE}$ by a factor of 0.01 in the training loss effectively giving more weight to the fully self-consistent datapoints. Furthermore, for \ch{CH} and \ch{OH}  $\lambda_n$ was scaled by 0.01. 

We augmented the G2/97 dataset with ground-state electron densities that we computed at the CCSD(T) level using the 6-311++G(3df,2pd) basis set, the same basis used for training the functionals.
Total atomic energies were taken from ref. \cite{chakravorty1993ground} and included in the training set as well. However, we disregarded energies for atoms that were not contained in the training data described above. Atomic electron densities were calculated and included for \ch{H} and \ch{Li}.

For model validation during training we used a disjoint subset of the data listed above, consisting of 8 atomization energies (\ch{C2H2}, \ch{BeH}, \ch{NO2}, \ch{S2}, \ch{CH4}, \ch{PF3}, \ch{CH2},\ch{C2H4O2}) and densities from G2/97, and two reference barrier heights from BH76 ( \ch{N2O} + \ch{H} $\rightarrow$ \ch{OH} + \ch{N2}, \ch{OH} + \ch{Cl} $\rightarrow$ \ch{O} + \ch{HCl}).

\subsection{Pretraining}

Models were pre-trained to match SCAN \cite{sun2015strongly}. Rather than training on atomization energies and densities, we make use of the fact that for known, semi-local functionals, the energy per unit particle can be computed exactly. This energy density, or correspondingly, the enhancement factor $F_{xc}$ can then be fitted on a grid-point level without the need for elaborate optimization procedures such as the self-consistent training used in this work. The functionals are pretrained on the 21 molecules contained in the training set by randomly sampling the exchange enhancement factor on molecular grids. For the exchange functional, we augmented this data by evaluating the enhancement factor on a regular grid in parameter space ($s$ and $\alpha$). Using this pretraining scheme initializes the models close to an optimal solution which ensures that the subsequent optimization converges within a reasonable amount of time.

\subsection{Initial density matrix}

The input density matrix supplied to SCF routine is a linear combination of an initial density matrix $\rho_{atomic}$ generated by PySCF from a combination of atomic contributions and a converged density matrix $\rho_{DFT}$ obtained with the pretraining model (SCAN)
\begin{align*}
    \rho_{init} &= (1 - \beta) \rho_{atomic} + \beta \rho_{DFT}\\
    \beta &= \frac{1}{2}(r+1)
\end{align*}
where $r$ is sampled from the uniform distribution $U(0,1)$ at every optimization step. This ensures that SCF calculations converge independently of the starting conditions.
We solve the KS equations using 25 SCF iterations, with linear density matrix mixing according to 
\begin{align*}
    \rho_{in, i+1} &= \alpha_{i} \rho_{out,i} + (1-\alpha_{i}) \rho_{in,i} \\
    \alpha &= 0.3^i + 0.3
\end{align*}
where $\rho_{out, i}$ is obtained by solving the KS equations at iteration $i$ given $\rho_{in,i}$. Linear density mixing is routinely used in DFT and is often required to have calculations converge within a reasonable number of iterations. From a machine-learning perspective, linear mixing is equivalent to so-called skip connections in residual neural networks, which allow gradients to propagate through deep networks more efficiently.

\subsection{Diagonalization}

Solving the Kohn-Sham equations involves generalized eigenvalue problems of the form 

\begin{equation}
    \mathbf{FC} = \mathbf{SC}\epsilon,
\end{equation}
commonly known as the Roothaan equations. Here, $\mathbf{F}$ represents the Fock matrix, $\mathbf{S}$ the overlap matrix between atomic orbitals , $\mathbf{C}$ a matrix of coefficients and $\epsilon$ the orbital energies. 
As PyTorch lacks an explicit algorithm to solve generalized eigenvalue problems (at least, at the time of completion of this manuscript), we reduce the Roothan equations to a standard symmetric eigenvalue problem by applying a Cholesky decomposition to $\mathbf{S}$ ahead of time, so that 

\begin{equation}
    \mathbf{S} = \vec L \vec L^T
\end{equation}
and therefore
\begin{equation}
    \vec L^{-1} \vec F \vec L^{-T} (\vec L^T \vec C) = \epsilon (\vec L^T \vec C).
\end{equation}

The equations can than be solved with PyTorch-native diagonalization routines. 

Backpropagating through the diagonalization causes diverging gradients in the case of degenerate eigenvalues. We propose an ad-hoc solution to this issue by adding a small, random pertubation to the XC-potential $V_\text{xc} \rightarrow V_\text{xc} + V_\text{noise}$ where $V_\text{noise}$ is drawn form a half-normal distribution with standard deviation $10^{-8}$ and subsequently symmetrized.
\subsection{Validation}

To not hinder efficient training, validation is done in parallel. This means that at every training epoch the model with its current parameters is saved to disk. A separate process, which is continuously run in the background, loads the model and conducts self-consistent calculation across the validation set. Training and validation set sizes are chosen so that these calculations finish before a new training epoch passes. The validation loss, identical to the training loss presented in the main text except for $\lambda_E=0$ and $w_j = \delta_{j,25}$, is then recorded. A checkpoint of the model is created if the validation loss decreased compared to its previous lowest value. We stop training if the validation loss does not decrease for 15 consecutive epochs. It should be pointed out that validation does not require keeping track of gradients. We can therefore use the original, much cheaper, PySCF implementation (modified to use our functionals) which allows us to tackle larger molecules than during training.

\subsection{Alternative density error metrics}

We have confirmed that our findings presented in the main text are largely independent of the density error metric chosen. 

Apart from a metric based on the absolute density deviation used in the main text 
\begin{equation}
    \varepsilon_{|n|} = \mathbb{E} \left[\frac{1}{N_e}\int_{\vec{r}} |n^{(i)}(\vec r) - n^{(i)}_{ref}(\vec r)|\right]\label{eq:densityerror},
\end{equation}
we considered the squared deviation 
\begin{equation}
    \varepsilon_{n^2} = \mathbb{E} \left[ \left(\frac{\int_{\vec{r}} (n^{(i)} - n^{(i)}_{ref})^2}{\int_{\vec {r}} (n^{(i)})^2 + \int_{\vec {r}}(n^{(i)}_{ref})^2} \right) ^{1/2}\right]\label{eq:densityerror},
\end{equation}
along with a squared deviation using an alternative normalization 
\begin{equation}
    \varepsilon_{\mathcal L} = \mathbb{E} \left[ \frac{1}{N_e}\left(\int_{\vec{r}} (n^{(i)} - n^{(i)}_{ref})^2 \right) ^{1/2}\right]\label{eq:densityerror}.
\end{equation}
Finally we looked at a density error defined in terms of the Kullback-Leibler divergence \cite{kullback1951information} or relative entropy between reference and predicted density:
\begin{equation}
    \varepsilon_{\text{KL}} = \mathbb{E} \left[ \frac{1}{N_e}\int_{\vec{r}} n^{(i)}_{ref}\log(\frac{n^{(i)}_{ref}}{n^{(i)}})\right]\label{eq:densityerror}.
\end{equation}

The $R^2$ values for the linear regression models relating density and energy errors are summarized in Tab. \ref{tab:r2}, density errors $\varepsilon$ as well as energy-density errors $\mathcal E \mathcal D$ are provided in Tab. \ref{tab:edalt}.

\begin{table}[!thb]
\begin{tabular}{c|cccc |c}
\toprule
Functional & s6 & a1 & s8 & a2 & cite Ref. \\
\hline
\textbf{xc-diff} & 1.000 & 0.493 & 0.501 & 4.459 & this work \\
SCAN \cite{sun2015strongly} & 1.000& 0.538 & 0.000& 5.420 & \cite{brandenburg2016benchmark}\\ 
TPSS \cite{tpss} &	1.000 &	0.454 &	1.944&	4.475 & \cite{goerigk2017look}\\
PBE \cite{pbe} &	1.000	& 0.429	& 0.788 &	4.441 & \cite{goerigk2017look} \\
revTPSS \cite{revTPSS} & 1.000 & 0.443    & 1.402 &	4.472 & \cite{goerigk2017look} \\
\hline
\end{tabular}
\caption{DFT-D3 dispersion correction coefficients.} \label{tab:d3}
\end{table}

\begin{table}[!thb]
    \centering
    \begin{tabular}{r|cccc}
    \toprule
          $R^2$ & $\varepsilon_{|n|}$ &   $\varepsilon_{n^2}$ &
         $\varepsilon_{\mathcal L}$ & $\varepsilon_{\text{KL}}$  \\
         \hline
          WTMAD-2 & 0.87 & 0.84 & 0.98 & 0.88\\
          \hline
    \end{tabular}
    \caption{$R^2$ correlation coefficients for the linear regression models relating density and energy errors for non-empirical functionals. Comparison of different density error metrics. }
    \label{tab:r2}
\end{table}

\begin{table}[!thb]
\begin{tabular}{r|ccc|ccc}
\toprule
{} &  AE &  BH & DE &  WTMAD-2 &  $\varepsilon_{|n|} \times 10^3$ &  $\mathcal E \mathcal D_{|n|}$ \\
\hline
PW91 \cite{perdew1992atoms}    &   15.6 &   9.9 &     20.3 &     11.3 &    10.8 &      11.5 \\
PBE \cite{pbe}    &   15.7 &   9.6 &     24.3 &     10.5 &    10.0 &      10.7 \\
RPBE  \cite{rpbe}  &    8.3 &   9.0 &     50.8 &     10.5 &     8.8 &      10.0 \\
B97  \cite{b97}   &    4.7 &   7.3 &     36.1 &      8.6 &     7.0 &       8.0 \\
OLYP \cite{olyp}   &    9.9 &   8.5 &     29.0 &      8.5 &    10.1 &       9.6 \\
revPBE \cite{revPBE} &    7.6 &   8.3 &     27.1 &      8.4 &     9.4 &       9.2 \\
\hline
MN12L \cite{mn12l}   &    4.2 &   1.7 &     20.9 &     11.3 &    11.1 &      11.6 \\
M11L \cite{m11l}   &    6.4 &   2.3 &     41.7 &      9.4 &    16.2 &      12.3 \\
TPSS \cite{tpss}   &    5.9 &   9.2 &     25.9 &      9.0 &     8.3 &       9.0 \\
M06L  \cite{m06l}  &    4.4 &   3.9 &     63.3 &      8.6 &     9.4 &       9.3 \\
revTPSS \cite{revTPSS}&    5.7 &   8.9 &     36.7 &      8.4 &     7.9 &       8.5 \\
SCAN   \cite{sun2015strongly} &    4.1 &   7.8 &     17.8 &      8.0 &     6.2 &       7.3 \\
\textbf{xc-diff} &    3.5 &   6.5 &     22.7 &      7.3 &     5.2 &       6.4 \\
\hline
HSE06 \cite{hse06}  &    3.6 &   4.6 &     14.3 &      6.6 &     5.6 &       6.3 \\
B3LYP  \cite{b3lyp} &    3.4 &   5.7 &     24.8 &      6.5 &     8.3 &       7.5 \\
PBE0  \cite{pbe0}  &    3.7 &   5.0 &     15.9 &      6.4 &     5.7 &       6.3 \\
M052X \cite{m052x}  &    4.0 &   1.7 &     26.3 &      4.6 &     7.5 &       5.8 \\
$\omega$B97X-V \cite{wb97xv} &    2.8 &   1.8 &     32.5 &      4.1 &     5.0 &       4.7 \\
\hline
\end{tabular}
\caption{Full list of functionals. Mean absolute errors in kcal mol$^{-1}$ for atomization energies (AE) over the W4-11 dataset, barrier heights (BH) in BH76 and decomposition energies (DE) for MB16-43. Weighted means WTMAD-2 and $\Delta$ are also given in kcal mol$^{-1}$. Mean density error $\varepsilon_n$ is unit-less.}
\end{table}

\begin{table}[!thb]
\begin{tabular}{r|cccc|cccc}
\toprule
{} &  $\varepsilon_{|n|}$ &   $\varepsilon_{n^2}$ & $\varepsilon_{\mathcal L}$ & $\varepsilon_{\text{KL}}$  &  $\mathcal E \mathcal D_{|n|}$ &   $\mathcal E \mathcal D_{n^2}$ & $\mathcal E \mathcal D_{\mathcal L}$ & $\mathcal E \mathcal D_{\text{KL}}$ \\
\hline
PW91    &    12.7 &  11.1 &      11.1 &   12.4 &          11.5 &      11.3 &            11.4 &         11.3 \\
PBE     &    11.8 &  10.2 &       9.8 &   10.9 &          10.7 &      10.5 &            10.3 &         10.2 \\
RPBE    &    10.3 &  11.6 &       9.8 &    7.9 &          10.0 &      11.2 &            10.4 &          8.6 \\
B97     &     8.2 &   8.1 &       8.1 &    8.9 &           8.0 &       8.5 &             8.5 &          8.4 \\
OLYP    &    11.9 &  11.9 &      10.4 &   11.8 &           9.6 &      10.0 &             9.5 &          9.5 \\
revPBE  &    11.2 &  10.9 &      10.2 &   11.7 &           9.2 &       9.6 &             9.4 &          9.4 \\
\hline
MN12L   &    13.1 &  16.2 &      17.2 &   14.0 &          11.6 &      13.4 &            13.8 &         12.0 \\
M11L    &    19.1 &  22.3 &      26.1 &   22.1 &          12.3 &      13.4 &            14.0 &         12.9 \\
TPSS    &     9.8 &   9.5 &       8.8 &    9.9 &           9.0 &       9.4 &             9.1 &          9.1 \\
M06L    &    11.1 &   9.7 &       8.7 &   14.1 &           9.3 &       9.2 &             8.8 &         10.3 \\
revTPSS &     9.4 &   8.6 &       8.1 &    9.5 &           8.5 &       8.6 &             8.4 &          8.5 \\
SCAN    &     7.3 &   6.7 &       7.8 &    9.6 &           7.3 &       7.4 &             8.0 &          8.4 \\
\textbf{xc-diff} &     6.2 &   5.1 &       4.9 &    3.8 &           6.4 &       6.1 &             6.0 &          4.7 \\
\hline
HSE06   &     6.6 &   5.1 &       5.7 &    6.2 &           6.3 &       5.9 &             6.3 &          6.1 \\
B3LYP   &     9.8 &   9.2 &       9.1 &    6.6 &           7.5 &       7.7 &             7.7 &          6.2 \\
PBE0    &     6.8 &   5.5 &       6.1 &    6.2 &           6.3 &       6.0 &             6.4 &          6.0 \\
M052X   &     8.9 &   8.1 &       9.2 &    8.3 &           5.8 &       5.9 &             6.2 &          5.7 \\
$\omega$B97X-V &     5.9 &  10.3 &       8.9 &    5.9 &           4.7 &       5.9 &             5.7 &          4.7 \\
\hline
\end{tabular}
\caption{Comparison of different density error metrics $\varepsilon$ and the corresponding energy-density errors $\mathcal E \mathcal D$ for all functionals considered in the main text. For better comparison, density errors $\varepsilon$ were normalized by their respective mean value and scaled by a factor of 10. $\mathcal{E} \mathcal{D}$ is provided in units of kcal mol$^{-1}$. Density errors are unit-less, except for $\epsilon_\mathcal L$ which has units of Bohr$^{-3/2}$}. \label{tab:edalt}
\end{table}
\end{document}